# Weyl Fermion Manipulation through Magnetic Transitions in the Ferromagnetic Non-Centrosymmetric Weyl semimetal PrAlSi

K. P. Wang[1,2*], W. J. Shi[1,3*†], W. Z. Cao[1,2*], X. T. Yang[1,2], Z. Y. Lv[4], C. Peng[5], C. Chen[1,2], D. F. Liu[6], H. F. Yang[1,2], L. X. Yang[7], M. Lyu[8], P. J. Sun[8], E. K. Liu[8], M. Ye[9,4], Y. L. Chen[1,2,3,5], Y. Sun[10], Y. P. Qi[1,2,11†] and Z. K. Liu[1,2†]

[1]*School of Physical Science and Technology, ShanghaiTech University, Shanghai 201210, China*

[2]*ShanghaiTech Laboratory for Topological Physics, ShanghaiTech University, Shanghai 201210, China*

[3]*Center for Transformative Science, and Shanghai High Repetition Rate XFEL and Extreme Light Facility (SHINE), ShanghaiTech University, Shanghai 201210, China*

[4]*Shanghai Institute of Microsystem and Information Technology, Chinese Academy of Sciences, Shanghai 200050, China*

[5]*Department of Physics, Clarendon Laboratory, University of Oxford, Parks Road, Oxford OX1 3PU, UK*

[6]*Department of Physics, Beijing Normal University, Beijing 100875, China*

[7]*State Key Laboratory of Low Dimensional Quantum Physics, Department of Physics, Tsinghua University, Beijing, 100084, China*

[8]*Beijing National Laboratory for Condensed Matter Physics, Institute of Physics, Chinese Academy of Sciences, Beijing 100190, China*

[9]*Shanghai Synchrotron Radiation Facility, Shanghai Advanced Research Institute, Chinese Academy of Sciences, Shanghai 201204, China*

[10]*Institute of Metal Research, Chinese Academy of Sciences, Shenyang, 110016, China*

[11]*Shanghai Key Laboratory of High-resolution Electron Microscopy, ShanghaiTech University, Shanghai 201210, China*

Email: shiwujun@shanghaitech.edu.cn, qiyp@shanghaitech.edu.cn, liuzhk@shanghaitech.edu.cn

**PrAlSi, a non-centrosymmetric ferromagnetic Weyl semimetal candidate with a Curie temperature of 17.8K, offers a unique platform for exploring the interplay of symmetry breaking and topological electronic structures. Up to now, the Weyl fermion distribution as well as their evolution across the ferromagnetic to paramagnetic phase transition in PrAlSi has not been explored. Here, we uncover the presence of Weyl fermions in PrAlSi and demonstrate they could be manipulated through the magnetic phase transition. Our ab-initio calculations indicate a shift in the momentum and energy positions of Weyl fermions, alongside an increase in Weyl point numbers due to band splitting. The predicted band splitting and shifting of Weyl fermions are corroborated by our angle-resolved photoemission spectroscopy experiments. Such manipulation of Weyl fermions leads to the appearance of a**

**net chirality charge and a significant modulation in optical conductivity, as proposed by our calculations. Our research presents a novel method for adjusting the properties of Weyl semimetals by controlling Weyl fermions through magnetic phase transitions, positioning PrAlSi as a model system.**

**INTRODUCTION**

In the rapidly developing field of topological materials, Weyl semimetals are distinguished by their topologically non-trivial electronic structures and multifaceted physical phenomena [1-5]. Characterized by linear electron dispersions (Weyl fermions) near nodal points known as Weyl points (WPs), Weyl semimetals feature these points as sources or sinks of Berry curvature in momentum space, each endowed with a specific chirality charge. Moreover, Fermi arc surface states connect a pair of WPs with opposite chirality [5-9]. The realization of Weyl semimetals hinges on the breaking of either inversion symmetry (IS) [5, 9-12] or time reversal symmetry (TRS) [7, 13-16], giving rise to a diverse array of materials. Weyl semimetals such as TaAs [2, 3, 5, 9-12, 17] do not possess IS but maintain TRS, whereas materials like $Co_3Sn_2S_2$ [14, 15, 18, 19] preserves IS while lacking TRS. Different crystal symmetry in Weyl semimetals leads to the different distribution of WPs and the chirality charges. The mirror symmetry (MS) and IS related WPs host opposite chirality charge while other symmetries-related WPs have the same chirality charge.

Weyl fermions and Fermi arcs in Weyl semimetals give rise to a variety of transport and optical phenomena, such as the chiral anomaly in magnetotransport [20, 21], pronounced second-harmonic generation (SHG) [22], and distinctive photocurrent effects [23]. Importantly, these characteristics can be modulated by altering the quantity, arrangement, and nature of WPs and Fermi arcs. To date, numerous strategies for this manipulation have been explored, including the application of external magnetic fields or pressure to transition Dirac semimetals to Weyl semimetals (Fig. 1a(i)) [24-26], enhancement of spin-orbit coupling (SOC) [17], and adjustment of WP distribution through external fields or pressure (Fig. 1a(ii)) [27-30], as well as the modification of Fermi arcs via surface dosing (Fig. 1a(iii)) [31]. Despite their effectiveness, these approaches often rely on external interventions like strain, magnetic ordering, element doping, or a vacuum environment, which may limit their practicality in device contexts. In this study, we introduce a new approach that leverages temperature-driven magnetic phase transitions, alongside TRS and MS breaking in noncentrosymmetric magnetic Weyl semimetals (Fig. 1a(iv)), exemplified by PrAlSi, to achieve Weyl fermion manipulation without the constraints of previous methods.

Recently, the noncentrosymmetric (thus IS is absent) RAlPn (R=rare earth, Pn=Si, Ge) compounds have been theoretically predicted and experimentally proved to be Weyl semimetals [32-52]. Notably, these materials exhibit diverse magnetic behaviors at low temperatures, ranging from antiferromagnetism in SmAlSi (Neel temperature $T_N$~11K) [38] and CeAlGe ($T_N$~5K) [35, 48, 53], to complex magnetic transitions in NdAlSi, which shows antiferromagnetic ordering at 7.2 K followed by ferromagnetism at 3.3 K [46, 50]. Among these non-centrosymmetric Weyl semimetal candidates, PrAlSi stands out with its ferromagnetic ordering at the highest high Curie temperature (Tc ~ 17.8 K) and the largest magnetic moment (3.6 μB/Pr) [39], positioning it as an ideal candidate to explore the interplay between magnetic phase transitions, symmetry breaking, and the manipulation of WPs (Fig. 1a(iv)). So far, although there have been studies [52, 54] on the effect of magnetism on WPs and Fermi arcs in RAlPn family. There have been reports [52] that the magnetic phase transition causes the surface state of CeAlSi to undergo splitting and the WPs to shift, while there is no change in the electronic structure during the magnetic phase transition in

PrAlSi and SmAlSi [54]. However, the existence of the Weyl fermions and their evolution through the ferromagnetic (FM) to paramagnetic (PM) phase transition, has not been systematically established in PrAlSi, especially by angle resolved photoemission spectroscopy (ARPES) measurements.

In this work, we carried out comprehensive ARPES experiments and *ab-initio* calculations to investigate the electronic structure of PrAlSi and its evolution from the PM phase to the FM phase. Our findings reveal a transition from 32 WPs in the PM phase to 56 in the FM phase, driven by band splitting due to the Zeeman effect. This band splitting and the WP shift were directly visualized in ARPES measurement. Furthermore, our theoretical calculation shows that the absence of inversion symmetry (IS) and mirror symmetry (MS) in the FM phase of PrAlSi allows for the manifestation of WPs with opposite chirality charges at different energy levels, introducing a net chirality charge. This unique arrangement of WPs can also introduce novel transport phenomena, such as the emergence of additional peaks in optical conductivity (OC) and large magneto-optic Kerr effect (MOKE) within the FM phase.

**RESULTS AND DISCUSSION**

*Structure, symmetry, and magnetic phase transition.* The PrAlSi has body-centered tetragonal structure in space-group 109 (I4$_1$md) (Fig. 1b). The crystal is characterized by a total of eight symmetry operations, which include two symmorphic symmetries (identity operation $E$, and mirror symmetry $m_y$) and six nonsymmorphic symmetries ($C_{2z}$, $C_{4z}^+$, $C_{4z}^-$, $m_x$, $m_{xy}$, and $m_{x\bar{y}}$). All constituent atoms occupy the 4*a* Wyckoff positions, with Pr at (0.5,0.5,0.0762), Si at the origin (0, 0, 0), and Al at (0,0,0.1654). Fig. 1c shows the crystal structure and the 3D BZ with the projected (001) surface BZ. Depending on the location in reciprocal space, the combined crystal symmetry and TRS ($\mathcal{T}$) can generate up to sixteen equivalent sites, although this multiplicity reduces to eight for certain positions, such as those within the $k_z = 0$ plane.

Due to the presence of the rare earth element Pr, PrAlSi exhibits ferromagnetic transition below 17.8 K with the magnetic momentum aligning along the *c* axis [39], as confirmed by our susceptibility measurements (Fig. 1d). Across the transition to the FM phase, symmetries such as $m_x$, $m_y$, $m_{xy}$, $m_{x\bar{y}}$, and $\mathcal{T}$ are broken, leaving only the $E$, $C_{2z}$ and $C_{4z}$ symmetries intact. Consequently, in the FM state, the crystal symmetry at most reciprocal-space points is reduced, generating only four equivalent sites, as opposed to the sixteen sites possible in the PM phase for most positions.

***Identification and manipulation of WPs.*** In our investigation of PrAlSi's electronic structure through first-principles calculations, we discerned 32 Weyl points (WPs) in the paramagnetic (PM) phase, constituting 16 WP pairs as cataloged in Table 1. The momentum distribution of these WPs within the first BZ is illustrated from both top (001) and side (010) views in Fig. S1. These WPs are categorized into three distinct types of inequivalent pairs: W$_1$, W$_2$, and W$_3$. Specifically, 8 W$_1$ pairs are situated in the $k_z = \pm 0.287 \text{Å}^{-1}$ plane, while both W$_2$ and W$_3$ types, comprising 4 pairs each, reside in the $k_z = 0$ plane (Fig. S1). Each pair consists of WPs of opposite chirality, denoted as $W_1^+/W_1^-$, $W_2^+/W_2^-$ and $W_3^+/W_3^-$ with '+' and '−' indicating positive and negative chirality charges, respectively. These pairs are symmetrically related by different mirror planes ($m_x$, $m_y$, $m_{xy}$, $m_{x\bar{y}}$), and intriguingly, they align at identical energy levels, as shown in Fig. 1f, ensuring a net chirality charge of zero below the Fermi level. However, by deliberately disrupting the MS by introducing magnetism, it becomes possible to shift the energy levels of WPs with opposing charges, as depicted in Fig. 1e. Such manipulation enables the induction of a net chirality charge beneath the Fermi level, opening avenues for novel topological phenomena.

As PrAlSi transitions to the FM phase at lower temperatures, MSs and TRS are disrupted, primarily due to magnetism from the Pr 4$f$ electrons. This leads to band splitting in the low-energy bands near the Fermi level, predominantly composed of Al and Si orbitals. This phenomenon, as evidenced by ARPES measurements, subsequently influences the manipulation of the WPs in the following aspects (locations listed in Table 1 and plotted in Fig. S1): (1) The emergence of additional WPs due to magnetism-induced band splitting, resulting in three more WP pairs near each $W_3$ pair and a total of 56 WPs in the FM phase. (2) Increment of momentum separation of existing WPs within the same group, driven by band splitting. Notably, $W_2$ and $W_3$ pairs experience significant shifts in momentum space. (3) In the FM phase, broken MSs lead to WPs of opposite chirality residing at different energy levels, enabling the adjustment of the Fermi level to achieve a net chirality charge (see detailed discussion in Supplemental Material). These manipulations underscore the intricate interplay between magnetic transitions and topological properties in PrAlSi.

***Calculations of the electronic structure.*** To elucidate the mechanisms behind the manipulation of WPs, we conducted electronic band structure calculations for PrAlSi along high-symmetry directions, considering both the presence and absence of SOC in the PM and FM phases (refer to Fig. S2). In the PM phase, spin-up and spin-down bands remain degenerate without SOC, while SOC induces splitting except at time-reversal invariant momentum (TRIM) points and along the $\Gamma - M$ line. At TRIM points, the TRS preserves the two-fold band degeneracy, whereas along the $\Gamma - M$ line, the $C_{4v}$ symmetry ensures it (Fig. S2a). Transitioning to the FM phase, magnetism disrupts TRS and MSs, leading to the splitting of spin-degenerate bands at TRIM points and along the $\Gamma - M$ line into distinct spin-up and spin-down bands, with SOC further widening this gap (Fig.

S2b).

The band splitting fosters the emergence and relocation of WPs, as detailed previously. Our calculations also extend to the Fermi surface for PrAlSi in both phases (Fig. S2c, d). In the PM phase, electron and hole pockets respect various symmetries, including $C_{2z}$, $C_{4z}^{+}$, $C_{4z}^{-}$, $m_x$, $m_{xy}$, and $m_{x\bar{y}}$, where as in the FM phase, only $C_{2z}$, $C_{4z}^{+}$, and $C_{4z}^{-}$ symmetries persist, with all pockets expanding in size, indicative of increased band crossings at the Fermi level.

Further analysis of the electronic band structures on the PrAlSi (001) surface, employing a slab model, reveals that the bulk states undergo splitting in the FM phase (Fig. 2d-f for the PM phase and Fig. 2j-l for the FM phase). We note the calculated surface states exhibit $C_2$ symmetry, diverging from the $C_4$ symmetry of bulk states. This discrepancy arises from the disruption of $C_{4z}$ symmetry due to the absence of translational symmetry along the *c*-axis upon cleaving the (001) surface.

***ARPES measurement of PrAlSi***. Our ARPES experiments on the (001) surface of PrAlSi were conducted in both PM and FM phases, as shown in Fig. 2. The measurements were conducted at photon energy of 94 eV, corresponding to $k_z$ =5.182 Å$^{-1}$ ($\frac{24\pi}{c}$), (see more $k_z$ measurement in Fig. S5). Constant energy contours (Fig. 2a, g) and dispersions along various high symmetry directions (Fig. 2b, c, h, i) highlight the presence of distinct bulk and surface band dispersions, closely matching our slab model predictions (Fig. 2d-f, j-l, also see Fig. S11), except for the surface states in FM state, which may be due to the weak coupling between f-electrons and the surface states or the surface reconstruction and contamination. The observation of relatively broad spectral weight distribution of bulk states indicates significant $k_z$ broadening (Fig. 2b, c, h, i, also see Fig. S5, S9). Our experimental results validate the existence of Weyl fermions, while the measured surface states exhibit $C_2$ symmetry (Fig. 2a, d, g, j), albeit slightly shifted in energy due to potential surface doping after cleavage. Although the electronic structure evolution across the FM to PM phase transition could not be clearly identified by the Fermi surface mapping in Fig. 2a, g, the temperature-dependent ARPES measurements along the $\overline{X}-\overline{\Gamma}-\overline{X}$ direction illustrate the band splitting and Weyl fermion ($W_3$) evolution across the FM to PM phase transition (Fig. 3).

Temperature-dependent ARPES measurements along the $\overline{X}-\overline{\Gamma}-\overline{X}$ direction illustrate the band splitting and Weyl fermion ($W_3$) evolution across the FM to PM phase transition (Fig. 3). Fig. 3a presents a set of spectra taken at consecutive temperatures, where the band splitting within the bulk valence band (BVB) in the FM state is observed, most prominent for the uppermost BVB near $\overline{\Gamma}$. This bulk band splitting is most pronounced at 8.9 K, decreases with temperature and disappears above Tc (Fig. 3d, also see the cooled downed data in Fig. S7). The magnitude of the splitting ($\Delta_{\text{gap}}$) follows a mean-field-like temperature dependence $\Delta \sim \sqrt{1-\frac{T}{T_c}}$, suggesting a direct correlation with magnetization (Fig. 3e). The band splitting along $\overline{\Sigma}-\overline{\Gamma}-\overline{\Sigma}$ and $\overline{X}-\overline{\Gamma}-\overline{X}$ with varying photon energies is documented in the supplemental materials (Figs. S6 a-d). These observations corroborate that band splitting occurs at TRIM points (see schematic in Fig. S6 e), as a result of TRS breaking, aligning with theoretical predictions.

Moreover, the ARPES data in the FM phase provides direct evidence of Weyl fermions' displacement associated with the band splitting. Detailed examination of the band dispersions near the Fermi level (Fig. 3b, c) revealed the emergence of additional band crossings in the FM phase, contrasted with the PM phase findings (indicated by the green and red arrows in Fig. 3b). This observation, aligned with slab model predictions, was substantiated by momentum distribution curve analysis (Fig.3c, f, h), which showed the merging of two bulk valence band peaks into a single peak across the FM to PM phase transition (Fig. 3f, h). A discernible band splitting was observed at low temperatures ($\Delta_{\text{split}} \sim 0.037 \text{ Å}^{-1}$ at 8.9K) which disappears above Tc, suggesting a band splitting induced Weyl fermion ($W_3$) shift. The momentum splitting is accompanied by a shift in the energy at the band top ($E_{\text{top}}$), as highlighted by the green arrow on the far left in Fig. 3b and c. This observation is corroborated by the analysis of the energy distribution curves (shown in Fig. 3g, h), which reveal that $E_{\text{top}}$ undergoes a shift of approximately 25meV towards the adjacent band across the transition from FM to PM states. Similar band shift in energy and momentum could also be observed in $W_1$ Weyl fermions (see Fig. S8). These results underscore the feasibility of manipulating Weyl fermions through band splitting and shifting, facilitated by magnetic phase transitions in PrAlSi.

*Optical conductivity and magneto-optical Kerr effect.* The transition from the PM to the FM phase, accompanied by the breaking of TRS and MS, sets the stage for the emergence of novel transport and optical phenomena, such as SHG, distinctive OC, and magneto-optical Kerr effect (MOKE) profiles. The OC, a critical parameter for probing allowed interband optical transitions, was rigorously analyzed in this study (Fig. 4a). The OC was computed using the Kubo formula [55, 56]:

$$\sigma_{\alpha\beta}(\omega) = -\frac{ie^2}{\hbar}\sum_{n,n'}\int\frac{d^3\boldsymbol{k}}{(2\pi)^3}\frac{f(n',\boldsymbol{k})-f(n,\boldsymbol{k})}{\varepsilon(n,\boldsymbol{k})-\varepsilon(n',\boldsymbol{k})}\times\frac{M^\alpha_{nn'}(\boldsymbol{k})M^\beta_{n'n}(\boldsymbol{k})}{\hbar\omega+\varepsilon(n',\boldsymbol{k})-\varepsilon(n,\boldsymbol{k})+i\eta}$$

where $\alpha,\beta = x,y,z$, $f = 1/[1+e^{(\varepsilon-\mu)/k_b T}]$ is the Fermi-Dirac distribution, $\mu$ is the chemical potential, $k_b$ is the Boltzmann constant, $T$ is the temperature, $M^\alpha_{nn'}(\boldsymbol{k}) = \langle n,\boldsymbol{k}|\hbar v_\alpha|n',\boldsymbol{k}\rangle = \left\langle\frac{\partial(n,\boldsymbol{k})}{\partial \boldsymbol{k}_\alpha}\Big|n',\boldsymbol{k}\right\rangle[\varepsilon(n,\boldsymbol{k})-\varepsilon(n',\boldsymbol{k})]$, $v_\alpha = \frac{1}{\hbar}\frac{\partial H(\boldsymbol{k})}{\partial \boldsymbol{k}_\alpha}$ is the velocity operator.

In the PM phase, TRS mandates that all off-diagonal OC tensor components vanish ($\sigma_{xy} = \sigma_{xz} = \sigma_{yx} = \sigma_{yz} = \sigma_{zx} = \sigma_{zy} = 0$), and the $C_{4z}$ crystal symmetry forces $\sigma_{xx} = \sigma_{yy}$. Conversely, in the FM phase, the magnetic moments along the z-axis disrupts TRS, eliminating $\sigma_{xz}$, $\sigma_{yz}$, $\sigma_{zx}$, and $\sigma_{zy}$ while enabling a nonzero $\sigma_{xy}$ and $\sigma_{yx}$ (anomalous Hall-like conductivity).

From the OC tensor, the complex Kerr angle $\phi_K = \theta_K + i\varepsilon_K$ of MOKE can be given by [57]

$$\phi_K(\omega) = \frac{-\sigma_{xy}}{\sigma_{xx}\sqrt{1 + i(4\pi/\omega)\sigma_{xx}}}$$

The calculated imaginary part OC in both PM ($\sigma_{xx}$ and $\sigma_{zz}$) and FM ($\sigma_{xy}$, $\sigma_{xx}$ and $\sigma_{zz}$) phase, and imaginary part Kerr angle are displayed in Fig. 4 c-d, e-h, and the included band are shown in Fig. 4a, b. The chemical potential ($\mu$) was fine-tuned within a $\pm 0.1$ eV range around the calculated Fermi level ($E_f$). The results underscore the anisotropy in OC, aligning with the crystal's symmetrical properties. Notably, the FM phase showcases additional OC peaks absent in the PM phase, attributed to photon-induced interband transitions among the four bands near the Fermi level. Among these, the anomalous Hall-like OC ($\sigma_{xy}$) and Kerr angle induced by the magnetic ordering in the FM phase, stands out as a pivotal finding (Fig. 4f). The largest Kerr angle is about 0.2 degree at small photon energies. This novel OC/MOKE behavior, particularly the photon-induced anomalous Hall conductivity, is anticipated to be experimentally observable by Terahertz spectroscopy, offering new insights into the optical properties of magnetic Weyl semimetals.

**CONCLUSION**

In conclusion, we systematically investigated the electronic structure of PrAlSi and confirmed it to be a Weyl semimetal in both PM and FM phase. Using the first-principles calculation and ARPES experiments, we uncover the manipulation of Weyl fermions through magnetic phase transition, including the number of WPs and the WP distribution in momentum and energy space. Specifically, this special distribution of WPs in the FM phase leads to net chirality charge beneath the Fermi level, and the emergence of new physical properties such as anomalous Hall-like OC and large Kerr angle.

Our results provide strong evidence for the influence between magnetism and topology in Weyl semimetals with both IS and TRS broken and provides new routes to manipulate Weyl fermions for spintronics and optoelectronics applications.

**METHODS**

*Crystal Growth:* Single crystals of PrAlSi were synthesized via the self-flux method [33, 39], employing high-purity Pr, Si, and Al in a 1:1:10 molar ratio. The mixture was placed in an alumina crucible within an argon atmosphere glove box, then vacuum-sealed in a quartz tube. The assembly was heated to 1100 °C over 24 hours, maintained at this temperature for 12 hours to ensure complete melting, and then gradually cooled to 750 °C at a rate of 2 °C/h. Excess Al was removed post-cooling using a centrifuge.

*ARPES experiment:* ARPES measurements were performed at BL 10.0.1, BL 7.0.2, BL 4.0.3 of the Advanced Light Source, BL 5-2 of the Stanford Synchrotron Radiation Light Source, alongside a helium lamp-equipped setup at Shanghaitech University. Measurements spanned temperatures from 5K to 250K under a vacuum below $5.0\times10^{-11}$ Torr. The setups offered an angular resolution of 0.1° and energy resolutions of better than 10 meV and 5 meV for photon energies of 100 eV and 21 eV, respectively.

*First-principles calculations:* The first-principles calculations were performed using the Vienna *ab initio* Simulation Package (VASP) [58]. The interactions between the valence electrons and ion cores are described by the projector augmented wave method [59, 60], and exchange-correlation potential was formulated by the generalized gradient approximation with the Perdew-Burke-Ernzerhof (PBE) scheme [61]. The Γ-centered 20×20×20 k-points were used for the first Brillouin-zone (BZ) sampling. The kinetic energy cutoff was set to 500 eV. Firstly, the crystal structure was fully relaxed until the force is less than 0.02 eV/Å, and the relaxed lattice constants of conventional cell are $a=b$=4.255 Å, and $c$=14.636 Å, which is close to the experiment results ($a=b$=4.23 Å, and $c$=14.55 Å). For the FM phase, the Pr 4$f$ electrons were treated as the valance electrons, while for the PM phase, the Pr 4$f$ electrons was treated as the core electrons. For the FM phase calculations, a Hubbard U=6 eV is applied on the Pr 4$f$ electrons to push the 4$f$ band away from the Fermi level, which are the origin of magnetic momentum. To calculate the surface states, a slab model of 10 conventional cells separated by 20 Å vacuum was employed. The SOC was included in calculations. For the OC calculations, a dense 60×60×60 k-points grid is employed. The chemical potential ($\mu$) was set ±0.1 eV around the calculated Fermi level ($E_f$), and a 0 to 0.1 eV range of photon energy was considered.

ACKNOWLEDGEMENT

Z. K. Liu acknowledges the support from the National Natural Science Foundation of China (92365204, 12274298) and the National Key R&D program of China (Grant No. 2022YFA1604400/03). W. J. Shi acknowledges the support from Science and Technology Commission of Shanghai Municipality (STCSM) (Grant No. 22ZR1441800), and HPC Platform of ShanghaiTech University. Y. P. Qi acknowledges the support from the National Natural Science Foundation of China (Grant No. 52272265) and the National Key R&D Program of China (Grant No. 2023YFA1607400).

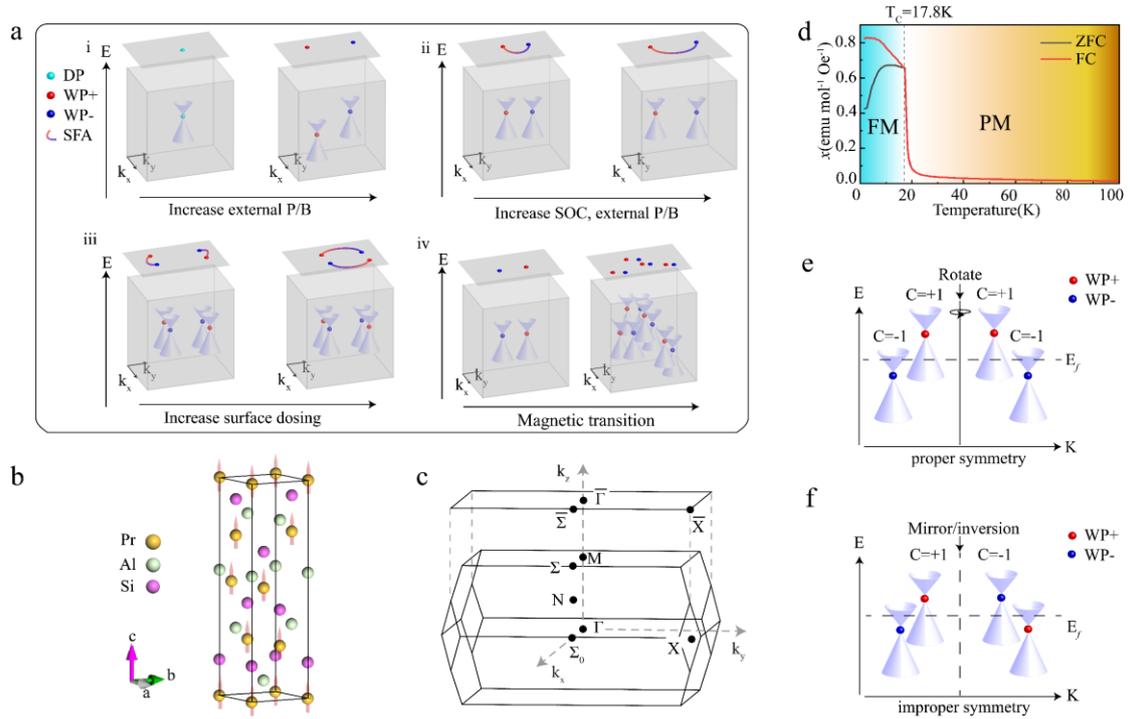

Fig. 1. Manipulations of Weyl Fermions (WPs) and properties of PrAlSi. (a) Possible scenarios of WPs and surface Fermi Arcs (SFA) manipulations. The red/blue dots represent WPs with opposite chirality charge. (i) Break degenerate Dirac point (DP) into WPs by external pressure (P) or magnetic field (B) through breaking IS or TRS. (ii) Enlarge the WPs' separation by increasing the SOC strength, or increasing external P/B. (iii) Change the connection of FAs by surface dosing. (iv)Manipulate and proliferate the WPs through TRS breaking magnetic transitions. (b) Crystal structure of PrAlSi. The magnetic order in FM phase is along the (001) direction, as indicated by the red arrows. (c) Illustration of the 3D BZ of PrAlSi, and the corresponding (001) surface BZ. (d) Zero-field cooled (ZFC, black curve) and field-cooled (FC, red curve, with magnetic field applied perpendicular to the $c$ axis) magnetic susceptibility ($\chi$) as functions of temperature. (e, f) Schematics of the proper (rotation) and improper (mirror and inversion) operation related WPs. $E_f$: Fermi level. The net chirality charge below $E_f$ is 2 in (e) and 0 in (f).

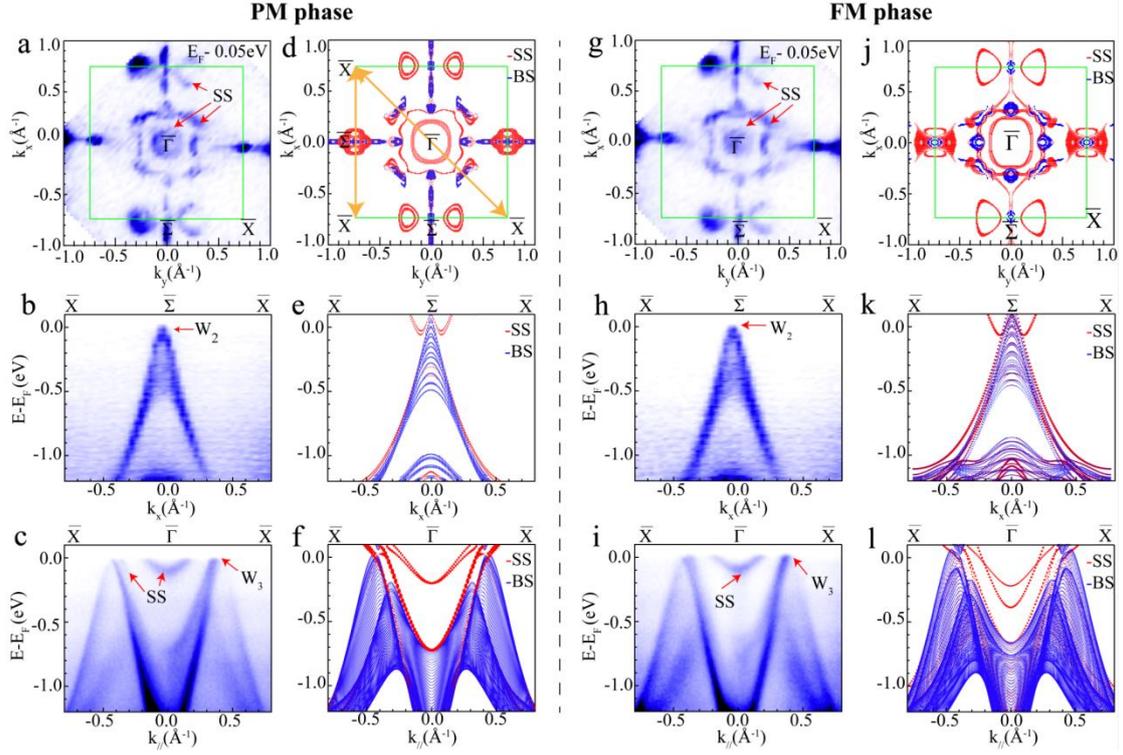

Fig. 2. Electronic band structure of PrAlSi. (a-f) The ARPES and slab calculation results of PrAlSi in the PM phase. (a) Intensity plot of the ARPES constant energy contour near $E_F$, integrated from $E_F$ to 0.05eV below. (b, c) Intensity plot of the band dispersions along the $\overline{X} - \overline{\Sigma} - \overline{X}$ (b) and $\overline{X} - \overline{\Gamma} - \overline{X}$ (c) directions as indicated by the arrows in (d). $W_2$ and $W_3$ label the positions of the WPs. (d-f) Corresponding slab calculations results. SS: surface electronic state. BS: bulk electronic state. (g-l) The same as (a-f) but in the FM phase.

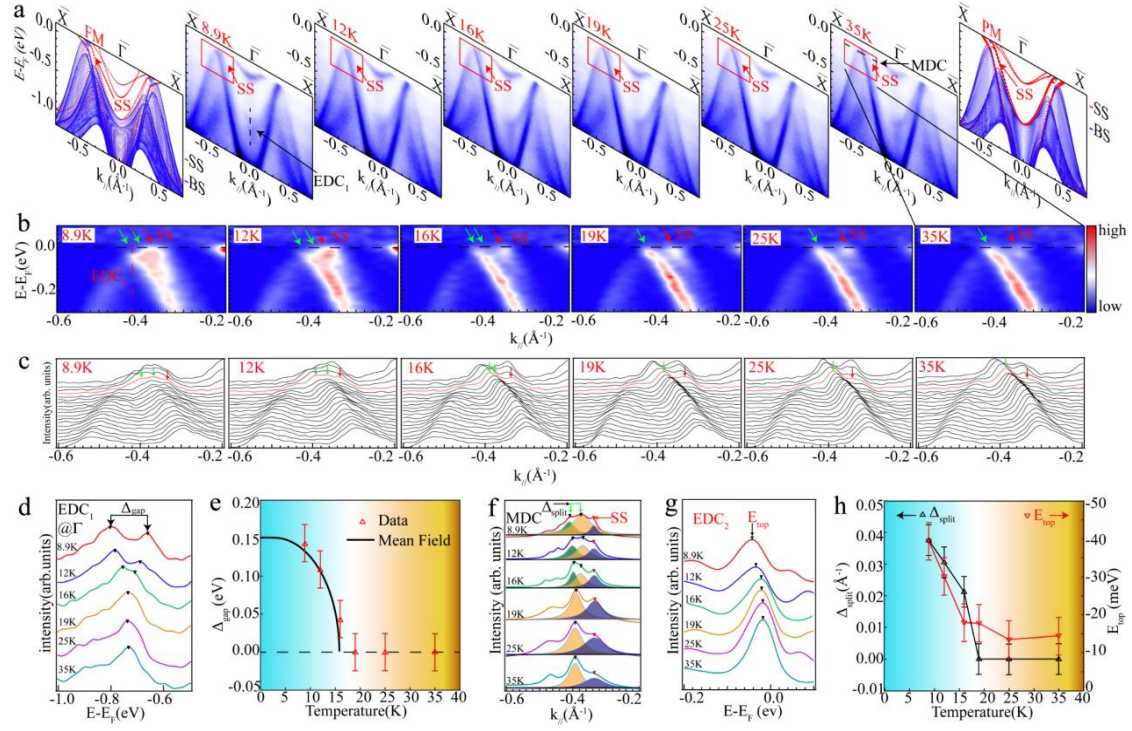

Fig. 3. Temperature evolution of the energy dispersion along $\overline{X}-\overline{\Gamma}-\overline{X}$ direction. (a) Intensity plots of the band dispersion along $\overline{X}-\overline{\Gamma}-\overline{X}$ direction at different temperatures. The calculated dispersions in the FM (PM) phase are labeled on the most left (right) panel. SS: surface electronic state. (b) The zoomed-in plots of the corresponding red rectangular boxes in (a). The green and red arrows label the bulk and surface bands, respectively. (c) The stacked line plots of the MDC of (b). (d) Plot of the EDCs at the $\overline{\Gamma}$ point at various temperatures. (e) Plot of the energy gap between two peaks marked by black arrows in (d) as a function of temperature. The black curve represents the mean-field fitting results. (f) Plot of the MDCs near $E_F$ (indicated by the black dashed line in the 35K data in (a) and red MDC in (c)) at different temperatures. The shaded peaks indicated the Lorentzian fitting results of the bulk (labelled by the green arrows) and surface (labelled by the red arrows) peaks, respectively. (g) Plot of the EDCs along the red dashed line (near the band top) in (b) at different temperatures. (h) Plot of the momentum splitting ($\Delta_{split}$) of the bulk peaks marked by green arrows in (f) and band top energy ($E_{top}$) marked in (g) as a function of temperature.

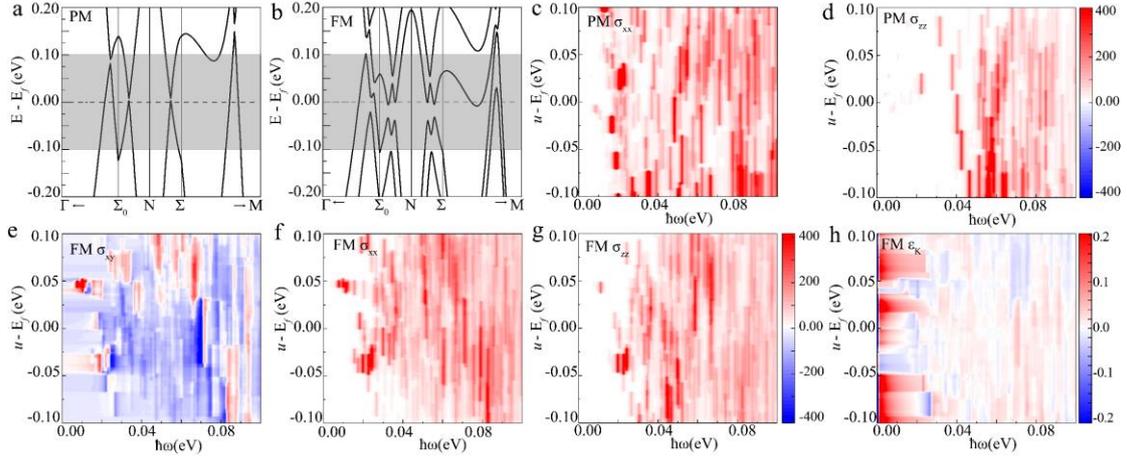

Fig. 4. The calculated optical conductivity and Kerr angles of PrAlSi. (a, b) The selected bands to calculate OCs and Kerr angles in the PM and FM phases, respectively. (c-d) The calculated imaginary part of OC $\sigma_{xx}$, and $\sigma_{zz}$ as a function of incident photon energy and the adjusted chemical potential. (e-g) The calculated imaginary part of OC $\sigma_{xy}$, $\sigma_{xx}$, and $\sigma_{zz}$ as a function of incident photon energy and the adjusted chemical potential. (h) The calculated imaginary part of Kerr angle $\varepsilon_K$. as a function of incident photon energy and the adjusted chemical potential. $\mu$: adjusted chemical potential. $E_f$: calculated Fermi level.

Table. 1. The designated groups of symmetry related WPs in PrAlSi. For each group, we list the number of WPs, and the momentum position and energy of one WPs.

| WPs | PM Phase | | | FM Phase | | |
|---|---|---|---|---|---|---|
| | Multiplicity | Coordinates (1/Å) | $E-E_f$ (meV) | Multiplicity | Coordinates (1/Å) | $E-E_f$ (meV) |
| $W_1^+$ | 8 | (0.01824, 0.37241, 0.28742) | 27.7 | 4 | (0.03767, 0.39084, 0.28062) | 28.9 |
| | | | | 4 | (0.05072, 0.35791, -0.28020) | 3.3 |
| $W_1^-$ | 8 | (0.37241, 0.01824, 0.28742) | 27.7 | 4 | (0.35774, 0.05037, 0.28062) | 3.6 |
| | | | | 4 | (0.39119, 0.03721, -0.28072) | 29.0 |
| $W_2^+$ | 4 | (0.74230, 0.00344, 0.0) | 61.9 | 4 | (0.76379, 0.01066, -0.03426) | 57.1 |
| $W_2^-$ | 4 | (0.00344, 0.74230, 0.0) | 61.9 | 4 | (0.01052, 0.76391, 0.03454) | 57.4 |
| $W_3^+$ | 4 | (0.26440, 0.34227, 0.0) | 91.6 | 4 | (0.26423, 0.42128, -0.00069) | 85.5 |
| | | | | 4 | (0.24377, 0.32376, -0.00361) | 54.1 |
| | | | | 4 | (0.30600, 0.25187, 0.05471) | 24.8 |
| | | | | 4 | (0.30602, 0.24967, -0.05556) | 29.3 |
| $W_3^-$ | 4 | (0.34227, 0.26440, 0.0) | 91.6 | 4 | (0.42145, 0.26411, 0.00068) | 85.6 |
| | | | | 4 | (0.32393, 0.24366, 0.00351) | 54.4 |
| | | | | 4 | (0.25192, 0.30588, -0.05469) | 24.7 |
| | | | | 4 | (0.24972, 0.30596, 0.05562) | 29.3 |